\documentclass[lettersize,journal]{IEEEtran}
\usepackage{amsmath,amsfonts}
\usepackage{graphicx,xcolor}
\usepackage{cite}
\usepackage{textcomp}
\usepackage{subcaption}
\usepackage{url}
\usepackage{listings}
\usepackage{cleveref}
\usepackage[T1]{fontenc}
\usepackage{balance}
\begin{document}

\title{A Constant-Time Implementation Methodology for Activation Functions on Microcontrollers}

\author{Andrii Tyvodar,
        Andreas Rechberger,
        Dirmanto Jap,
        Shivam Bhasin,
        Bernhard Jungk,
        Jakub Breier,
        and Xiaolu Hou%
\thanks{A. Tyvodar and X. Hou are with the Faculty of Informatics and Information Technologies,
Slovak University of Technology in Bratislava, Bratislava, Slovakia
(e-mail: xtyvodar@stuba.sk; xiaolu.hou@stuba.sk).}%
\thanks{X. Hou is also with the State Key Laboratory of Blockchain and Data Security,
Zhejiang University, Hangzhou, China.}%
\thanks{A. Rechberger is an independent researcher,
(e-mail: andreas@rechberger.li).}%
\thanks{D. Jap and S. Bhasin are with Temasek Laboratories, Nanyang Technological University,
Singapore (e-mail: djap@ntu.edu.sg; sbhasin@ntu.edu.sg).}%
\thanks{B. Jungk is with the Faculty of Computer Science, Albstadt-Sigmaringen University,
Albstadt, Germany (e-mail: jungk@hs-albsig.de).}%
\thanks{J. Breier is with TTControl GmbH, Vienna, Austria
(e-mail: jbreier@jbreier.com).}%
}

\maketitle

\begin{abstract}
Embedded neural-network inference can leak information through timing side channels, including leakage caused by the evaluation of activation functions. 
This work proposes a constant-time implementation methodology for activation functions on embedded microcontrollers and validates it on ReLU, sigmoid, tanh, GELU, and Swish on an ARM Cortex-M4 platform.
The proposed methodology combines branchless selection, fixed-cost Pad\'e-based approximation, dummy arithmetic where needed, and cycle alignment to obtain timing-regular activation-function implementations.
As motivation, we also evaluate a desynchronization-based countermeasure and show that it remains vulnerable to a template-based timing attack. 
Experimental results show that the resulting protected implementations achieve identical cycle counts for all tested inputs, including \(88\) cycles in the three-function setting and \(108\) cycles in the five-function setting.
At the same time, the numerical-error analysis indicates that the approximated nonlinear functions retain high accuracy.
These results suggest that the proposed methodology provides a practical basis for constructing side-channel-resistant activation functions in embedded inference.
\end{abstract}


\section{Introduction}

Deep neural networks (DNNs) are increasingly deployed on embedded and edge platforms, where inference must be executed under strict constraints on latency, memory usage, and energy consumption. 
In such settings, software implementations are often optimized at a low level in order to achieve practical performance on microcontrollers and other resource-constrained devices. 
Prior work has studied neural-network deployment through hardware/software co-exploration~\cite{jiang2020hardware}, timing-constrained implementation~\cite{jiang2018heterogeneous}, and timing-aware scheduling of DNN workloads~\cite{kang2024batch}. 
At the same time, these implementations may expose side-channel leakage, including timing-dependent behavior that can reveal information about the internal computation~\cite{batina2022implementation}.

Recent work has shown that neural-network implementations are vulnerable to side-channel analysis, enabling reverse engineering of model architectures and extraction of information about intermediate computations and processed inputs~\cite{batina2019csi,horvath2024sok,maji2021leaky}.
In particular, activation functions have been identified as a source of distinguishable leakage, since different nonlinearities and different input regions may exhibit characteristic execution patterns~\cite{takatoi2020simple,batina2019csi}. 
This is especially relevant in embedded inference, where timing observations may be accessible to an attacker and where even seemingly simple implementation choices can affect side-channel behavior.

A recent direction for mitigating such leakage is to apply hiding-oriented countermeasures, such as desynchronization, in which random delays are introduced in order to obscure the timing pattern of activation-function evaluation~\cite{breier2023desynchronization}. 
Such techniques can make the leakage harder to interpret, but they do not remove the underlying input-dependent timing behavior itself. 
This raises the question of whether a stronger mitigation can be achieved by redesigning activation-function implementations so that their execution time no longer depends on the processed input.

\begin{figure*}[!tb]
    \centering
    \includegraphics[width=0.8\linewidth]{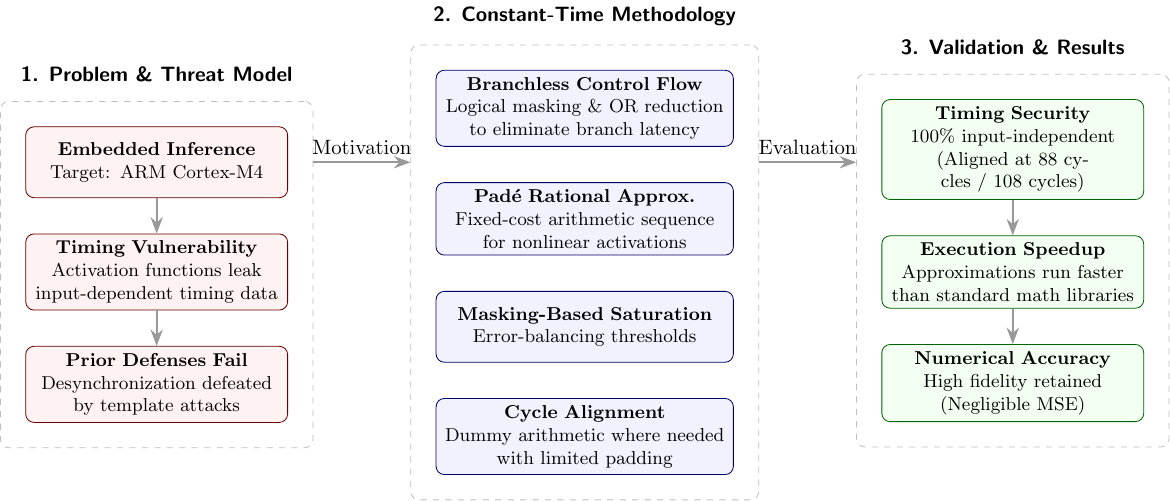}
  \caption{Summary of the threat model, proposed methodology, and experimental results. 
  Timing leakage in embedded activation-function evaluation is mitigated using a combination of masking-based selection, Pad\'e approximation, fixed-cost dummy arithmetic, and limited cycle-alignment padding. 
  This supports uniform execution latency for both nonlinear and piecewise-linear activation functions while maintaining low approximation error for the nonlinear functions.}
    \label{fig:overview}
\end{figure*}

In this work, we propose a constant-time implementation methodology for activation functions and instantiate it on five widely used examples: ReLU, sigmoid, \(\tanh\), Gaussian Error Linear Unit (GELU)~\cite{hendrycks2016gaussian}, and Swish~\cite{ramachandran2017searching}.
The proposed methodology combines branchless, masking-based selection, cycle-aligned implementation techniques, and, for nonlinear activations, Pad\'e-type rational approximation in order to construct implementations whose execution time is independent of the input value. 
In contrast to prior work that uses approximation primarily for efficiency, here the approximation also serves as a component of the side-channel countermeasure itself.
The overview of the proposed approach, along with the threat model and the experimental results is provided in Fig.~\ref{fig:overview}.

The contributions of this work are as follows:
\begin{itemize}
    \item We propose a constant-time implementation methodology for activation functions on embedded microcontrollers, based on branchless, masking-based selection, fixed-cost Pad\'e approximation based computation, and cycle alignment.
    \item We instantiate this methodology on ReLU, sigmoid, \(\tanh\), GELU, and Swish, using a shared rational approximation for the nonlinear components.
    \item We show, for ReLU, sigmoid, and \(\tanh\), that a desynchronization-based countermeasure can still be attacked using a template-based timing analysis.
    \item We demonstrate experimentally that the protected implementations exhibit identical cycle counts for all tested inputs, including a five-function evaluation over the extended interval \([-500,500]\) in which all protected implementations execute in \(108\) clock cycles.
    \item We evaluate the resulting trade-offs in execution time and numerical accuracy.
\end{itemize}

The remainder of this paper is organized as follows. 
Section~\ref{sec:threat_model} introduces the considered threat model. 
Section~\ref{sec:related_work} reviews the most relevant prior work. 
Section~\ref{sec:ct_design} presents the proposed constant-time implementation methodology and its instantiation on the considered activation functions. Section~\ref{sec:experimental_setup} describes the experimental setup. 
Section~\ref{sec:results} presents the experimental results, Section~\ref{sec:discussion} discusses their implications and limitations, and Section~\ref{sec:conclusion} concludes the paper.

\section{Threat Model}
\label{sec:threat_model}

We consider an attacker who targets embedded neural-network inference implemented on a microcontroller. 
The attacker is assumed to be able to observe the execution time of activation-function evaluations, either directly through local interaction or indirectly through repeated timing measurements of the device. 
In the profiled setting considered in this work, the attacker may additionally collect reference measurements from known activation functions to build statistical models and use them to classify unknown observations.

The attacker's objective is to exploit timing leakage to distinguish activation functions and, more generally, to infer information about the internal behavior of the neural-network implementation. 
This information can support higher-level goals such as model analysis, reverse engineering, or the identification of sensitive computation patterns. 

The goal of the proposed methodology is therefore to suppress timing leakage in activation-function execution, rather than merely obscure it through randomization or desynchronization.

\section{Background and Related Work}
\label{sec:related_work}

This section reviews prior work most relevant to the present study. 
Section~\ref{subsec:sca_nn} summarizes side-channel attacks on neural-network implementations, with emphasis on leakage from activation functions. Section~\ref{subsec:timing_countermeasures} discusses countermeasures against timing-based side-channel attacks, including hiding-oriented approaches. Section~\ref{subsec:constant_time_related} then positions the present work with respect to constant-time implementation techniques, while Section~\ref{subsec:activation_approx_related} reviews approximation-based activation-function implementations for embedded inference.

\subsection{Side-Channel Analysis Attacks on Neural-Network Implementations}
\label{subsec:sca_nn}
Side-channel analysis (SCA) attacks on neural-network implementations have received increasing attention in recent years. 
Prior work has shown that physical leakages such as electromagnetic emanations, power consumption, and timing behavior can reveal information about model architectures, secret parameters, and confidential input data~\cite{batina2019csi,horvath2024sok,yu2020deepem,maji2021leaky,wu2025catch}.
Among the components of neural-network inference, activation functions have been identified as a particularly informative source of leakage. 
\cite{takatoi2020simple} demonstrated that activation functions exhibit distinct EM emanation patterns and can be distinguished accordingly.
\cite{batina2019csi} showed that the activation type can be inferred from execution-time behavior.

More generally, timing-based side-channel attacks exploit input-dependent execution latency to infer sensitive information about the target system. 
In the context of machine learning, such attacks have been shown to reveal information about inputs, classes, and model behavior~\cite{shukla2023whispering,akinsanya2024timing}.

\subsection{Desynchronization and Hiding-Based Countermeasures Against SCA Attacks}
\label{subsec:timing_countermeasures}

Compared with the growing literature on side-channel attacks against neural-network inference, work on dedicated countermeasures against timing-based SCA attacks remains relatively limited. 
To the best of our knowledge, the only existing countermeasure against timing side-channel attacks on activation functions was proposed by~\cite{breier2023desynchronization}.
In this work, the authors studied a desynchronization-based countermeasure for activation-function implementations, in which random delays are inserted so that the timing dependence on both the input and the activation type becomes harder to exploit. 
This line of work falls into the broad concept of hiding-based countermeasures against SCA.

More broadly, hiding-based countermeasures~\cite{mangard2007power} such as random delay insertion, jitter, and operation shuffling are well established in side-channel protection for embedded systems, but mostly against power/EM-based SCA attacks. 
\cite{puvskavc2025make} proposed to shuffle multiplication computations for each neuron to counter correlation power analysis.
\cite{yan2023defense} uses ring-oscillator-based fine-grained noise generation together with an algorithm-level obfuscation step to protect FPGA DNN accelerators against power SCAs.
\cite{narkthong2025permutev} uses random permutation of loop-iteration execution order in hardware to obfuscate the EM signature of NN inference on a RISC-V core.

Section~\ref{subsec:attack} shows that template-based attacks~\cite{hou2024cryptography} remain effective against the evaluated desynchronization-style countermeasure.
In contrast to hiding-oriented approaches, the present work explores a constant-time design strategy for activation functions, with the aim of eliminating input-dependent timing variation at the implementation level. 
The proposed approach is therefore complementary to prior work on desynchronization-based defenses and can be viewed as a more direct mitigation of timing leakage in activation-function evaluation.

\subsection{Constant-Time Implementation as a Countermeasure}
\label{subsec:constant_time_related}
Constant-time implementation is a well-established countermeasure against timing side-channel attacks on cryptographic implementations~\cite{almeida2016verifying}, where the objective is to ensure that the execution path and runtime do not depend on secret or sensitive data. 
In contrast to hiding-based approaches such as random delays or desynchronization, constant-time design aims to suppress the leakage source itself by avoiding data-dependent branches, memory accesses, and other input-dependent execution effects.

Despite its importance in secure software design, the constant-time paradigm has received comparatively little attention in the context of neural-network inference. 
The present work adopts the constant-time perspective specifically for activation functions. 
In this setting, the goal is not only to flatten the timing profile empirically, but to provide a general implementation methodology by which activation functions can be constructed so that the executed instruction sequence is independent of the input value.

\subsection{Approximation of Activation Functions for Embedded Inference}
\label{subsec:activation_approx_related}
Approximation of nonlinear activation functions is a common technique for reducing the computational cost of neural-network inference on resource-constrained platforms. 
In particular, functions such as sigmoid and \(\tanh\) are often replaced by polynomial, rational, or piecewise approximations in order to avoid the expense of direct evaluation through standard transcendental functions~\cite{hush1998efficient,liu2023cost,liu2025dif}. 
Prior work has primarily studied such approximations from the perspective of computational efficiency and numerical accuracy in embedded and low-power inference settings. 
In this literature, the main objective is typically to reduce execution time, code complexity, or resource usage while preserving acceptable approximation quality. 

The present work is related to this line of research in that it also employs a compact approximation of nonlinear activation functions. 
However, the goal here is different: the approximation is used not only to improve efficiency, but also to enable constant-time execution. 
The shared rational approximation adopted in this work supports a fixed sequence of arithmetic operations together with branchless saturation logic, making it suitable for timing-regular software implementation on the target microcontroller. 
In this sense, the approximation is part of a broader constant-time implementation methodology rather than merely a means of accelerating inference.

\section{Constant-Time Implementation Methodology}
\label{sec:ct_design}

This section presents the proposed constant-time implementation methodology for activation functions. 
The objective is to provide a reusable construction principle by which activation functions can be implemented with input-independent execution behavior while maintaining acceptable numerical accuracy for inference-oriented workloads. 
Section~\ref{subsec:activation_definitions} first defines the activation functions considered in this work and distinguishes the reference functions from their protected approximated implementations. 
Section~\ref{sec:strategy} then introduces the constant-time design strategy based on branchless, masking-based selection. 
Sections~\ref{sec:approximation}--\ref{subsec:threshold} describe the shared \(R_{\tanh}\)-based approximation, masking-based saturation, and saturation-threshold selection. 
Finally, Sections~\ref{subsec:relu_instantiation} and~\ref{subsec:implementation_considerations} discuss the special treatment of ReLU and the implementation measures used to preserve the intended timing-regular behavior after compilation.

\subsection{Activation Functions Considered}
\label{subsec:activation_definitions}

This work considers five activation functions commonly used in neural-network inference: ReLU, sigmoid, \(\tanh\), GELU, and Swish. 
Their standard mathematical definitions are summarized here in order to distinguish the reference functions from the protected approximated implementations introduced later.

The rectified linear unit is defined as
\begin{equation}
\label{eq:relu_def}
\mathrm{ReLU}(x)=\max(0,x).
\end{equation}

The sigmoid function is defined as
\begin{equation}
\label{eq:sigmoid_def}
S(x)=\frac{1}{1+e^{-x}}.
\end{equation}

The hyperbolic tangent is defined as
\begin{equation}
\label{eq:tanh_def}
\tanh(x)=\frac{e^x-e^{-x}}{e^x+e^{-x}}.
\end{equation}

The Gaussian Error Linear Unit is defined as
\begin{equation}
\label{eq:gelu_def}
\mathrm{GELU}(x)
=
x\Phi(x)
=
\frac{x}{2}
\left(
1+\mathrm{erf}\left(\frac{x}{\sqrt{2}}\right)
\right),
\end{equation}
where \(\Phi(\cdot)\) denotes the cumulative distribution function of the standard normal distribution.

The Swish activation is defined as
\begin{equation}
\label{eq:swish_def}
\mathrm{Swish}_{\beta}(x)=xS(\beta x).
\end{equation}
In this work, we use the common choice \(\beta=1\), so that
\begin{equation}
\label{eq:swish_beta_one_def}
\mathrm{Swish}(x)=xS(x).
\end{equation}

\subsection{Constant-Time Design Strategy}
\label{sec:strategy}
A central principle of the proposed methodology is that the execution path must not depend on the input value.
Conventional activation-function implementations may contain data-dependent branches, early exits in saturation regions, or piecewise approximations with input-dependent interval selection, all of which can introduce observable timing differences.

% A simplified example of such runtime dependence can be illustrated using the ReLU function.
% It either does nothing or replaces the value with zero. 
% This operation consists of two fundamental steps. First, a (usually arithmetic) operation, which creates a boolean value, followed by a conditional branch to execute different sequences based on that result.
% Only the latter of which is contributing to the run time dependency. Due to cache (or flash prefetch) aspects, the branch instruction as such may differ in timing whether the branch is taken or not. In addition, the sequences of the two conditions may be different. In the example of the ReLU either assign a value of zero to a variable or do nothing.

A simplified example of such runtime dependence can be illustrated using the ReLU function.
A conventional implementation either leaves the value unchanged or replaces it with zero. 
This operation typically consists of two fundamental steps: first, a comparison operation creates a Boolean value, and second, a conditional branch selects the instruction sequence to execute based on that result.
The latter can contribute to runtime dependence. Due to cache or flash-prefetch effects, the branch instruction itself may differ in timing depending on whether the branch is taken or not. In addition, the two branch paths may contain different instruction sequences. In the ReLU example, one path may assign zero to a variable, while the other may leave the value unchanged.

% To avoid these effects, each protected activation function is evaluated using a fixed sequence of arithmetic operations. 
% For conditionally selecting one out of two values (like ReLU or saturation) we sign extend the boolean to an all zero or all one mask (0xFFFF, 0x0000 in the 16 bit case) and perform a masking of the two alternatives, followed by logical OR reduction.
To avoid these effects, each protected activation function is evaluated using a fixed-cost sequence of operations.
For conditionally selecting one out of two values, several strategies exist. For floating-point values, a possible coding is shown in \cref{code:ct_sel_f32}. The implementation derives a 32-bit all-zero or all-one mask from the comparison result and applies this mask to the IEEE-754 bit representations of the two candidate values, followed by a bit-wise OR combination.
As this is a sequence identical for any input value, this prevents the control flow impact, effectively implementing a branch-less conditional. The executed instruction sequence remains identical across different input values, making the approximation well suited to constant-time execution.

\lstset{
	language=C++,                % choose the language of the code
	stepnumber=1,                   % the step between two line-numbers.        
	numbersep=5pt,                  % how far the line-numbers are from the code
	backgroundcolor=\color{white},  % choose the background color. You must add \usepackage{color}
	showspaces=false,               % show spaces adding particular underscores
	showstringspaces=false,         % underline spaces within strings
	showtabs=false,                 % show tabs within strings adding particular underscores
	tabsize=2,                      % sets default tabsize to 2 spaces
	captionpos=b,                   % sets the caption-position to bottom
	breaklines=true,                % sets automatic line breaking
	breakatwhitespace=true,         % sets if automatic breaks should only happen at whitespace
	title=\lstname,                 % show the filename of files included with \lstinputlisting;
	basicstyle=\small,
	frame=single,
	literate={~}{{\fontfamily{ptm}\selectfont \textasciitilde}}1,
}
\begin{lstlisting}[caption={Branchless conditional select for 32-bit floating-point values.}, label=code:ct_sel_f32]
float ct_select_f32(float a, float b, uint32_t mask) {
	uint32_t ua = std::bit_cast<uint32_t>(a);
	uint32_t ub = std::bit_cast<uint32_t>(b);
	uint32_t ur;
	ur = (ua & ~mask) | (ub & mask);
	return std::bit_cast<float>(ur);
}
\end{lstlisting}

In the present work, this methodology is instantiated on ReLU, sigmoid, \(\tanh\), GELU, and Swish, but its components are not limited to these five cases.

Although the ReLU implementation uses the conventional ternary expression 
\texttt{return (x > 0) ? x : 0}, this expression was compiled into a timing-regular 
IT-instruction-based sequence on the target platform. In contrast, applying the same 
vanilla conditional-selection style to the saturation paths of the nonlinear functions 
did not result in timing-regular behavior in our experiments. Therefore, for 
sigmoid, \(\tanh\), GELU, and Swish, we used an explicit bit-mask-based selection 
to avoid relying on compiler-dependent behavior regarding conditional expressions.

This methodology can be implemented in a high-level language 
(C or C++) without resorting to inline assembly. 
Implementing any function in a high-level language with the goal of achieving constant timing is subject to the effects of compiler optimization.
As such, careful testing is required. 
However, assembly implementations also suffer from similar effects, caused by microarchitectural details, such as memory alignment, cache effects, or branch prediction, which also affect the timing. 
Hence, a one to one mapping from assembly instructions to cycles is no longer a trivial task. For this work, it was assumed that using C++ provides more benefits than additional burden.

The nonlinear implementations, namely sigmoid, \(\tanh\), GELU, and Swish, were verified to retain their intended 
branchless structure under the tested optimized builds (\texttt{-O1}, \texttt{-O2}, and
\texttt{-O3}), when using the explicit bit-mask-based selection. For ReLU, the 
conventional ternary expression was retained, since the generated code for the 
optimized build, used in the measurements, was timing-regular on the target platform as shown in \cref{code:ReLU_asm}. 
A debug build (\texttt{-O0}) is unsuitable because its debug-oriented code generation does not 
reliably preserve side-channel-resistant binaries. In a larger project, compiler 
optimizations must be chosen carefully to retain this structure. In particular, 
whole-program link-time optimization may break the intended branchless form.

The methodology is evaluated for finite floating-point inputs in the specified experimental domains. Special IEEE-754 values such as NaNs and infinities are outside the considered input model.

\subsection{Approximation Component of the Methodology}
\label{sec:approximation}

For nonlinear activation functions, the proposed methodology employs a Pad\'e-type rational approximation~\cite{george1996pade} as a fixed-cost computational component. 
In the present instantiation, this approximation is defined for \(\tanh\) as
\begin{equation}
\label{eq:core}
R_{\tanh}(x) = x \cdot \frac{P(x^2)}{Q(x^2)},
\end{equation}
where \(P(\cdot)\) and \(Q(\cdot)\) are low-degree polynomials in \(x^2\). 
Specifically, \(\tanh\) is approximated as
\[
\tanh(x)\approx R_{\tanh}(x)
=
x
\frac{1 + \dfrac{5}{39}x^{2} + \dfrac{2}{715}x^{4} + \dfrac{1}{135135}x^{6}}
     {1 + \dfrac{6}{13}x^{2} + \dfrac{10}{429}x^{4} + \dfrac{4}{19305}x^{6}}.
\]
This form requires only a fixed sequence of additions, multiplications, and one division, without data-dependent loop bounds or input-dependent interval selection. 
In addition, the use of \(x^2\) preserves odd symmetry, which is consistent with the behavior of \(\tanh(x)\).

The same \(R_{\tanh}\)-based computation is then reused to construct the other nonlinear activation functions considered in this work. 
For sigmoid, denoted by \(S\), we use the standard identity
\[
S(x) = \frac{1}{2}\left(1 + \tanh\left(\frac{x}{2}\right)\right),
\]
which yields the approximation
\begin{equation}
\label{eq:sigmoid_core}
S(x)
\approx
\frac{1}{2}
+
\frac{1}{2}\,R_{\tanh}\!\left(\frac{x}{2}\right).
\end{equation}

Swish is defined in this implementation as
\[
\mathrm{Swish}(x) = x \cdot S(x),
\]
corresponding to the common choice \(\beta=1\). 
Using Equation~(\ref{eq:sigmoid_core}), the protected Swish implementation is therefore based on
\begin{equation}
\label{eq:swish_core}
\mathrm{Swish}(x)
\approx
x\left(
\frac{1}{2}
+
\frac{1}{2}R_{\tanh}\!\left(\frac{x}{2}\right)
\right).
\end{equation}

For GELU, we use the common \(\tanh\)-based approximation
\[
\mathrm{GELU}(x) \approx \frac{x}{2}\left(1 + \tanh\left(\sqrt{\frac{2}{\pi}}\left(x + 0.044715x^3\right)\right)\right).
\]
Replacing \(\tanh\) with \(R_{\tanh}\) gives the implementation form
\begin{equation}
\label{eq:gelu_core}
\mathrm{GELU}(x)
\approx
\frac{x}{2}
\left(
1+
R_{\tanh}\!\left(
\sqrt{\frac{2}{\pi}}\left(x+0.044715x^3\right)
\right)
\right).
\end{equation}

Equations~(\ref{eq:core})--(\ref{eq:gelu_core}) provide fixed-operation-count arithmetic expressions for the nonlinear activations.
The saturation logic described in Section~\ref{sec:saturation} is then applied without data-dependent control flow in order to preserve timing-regular execution outside the approximation region.

\subsection{Masking-Based Saturation in the Proposed Methodology}
\label{sec:saturation}

Within the proposed methodology, fixed-cost \(R_{\tanh}\)-based computation is combined with explicit saturation in order to obtain well-behaved numerical properties while preserving timing-regular execution. 
For large-magnitude finite inputs, the nonlinear activation functions approach simple limiting forms: \(\tanh(x)\) approaches \(\pm 1\), sigmoid approaches either \(0\) or \(1\), and both GELU and Swish approach \(0\) for large negative inputs and \(x\) for large positive inputs.
The implementation therefore uses saturation outside the selected approximation intervals.

To preserve constant-time behavior, this saturation is implemented without conventional branching. 
Instead, the approximation result and the saturated output candidates are represented in arithmetic form, and the final output is selected using the masking-based selection strategy described in Section~\ref{sec:strategy}. 
When required, the input passed to the \(R_{\tanh}\)-based computation is also clamped to the selected approximation interval using the same masking-based strategy. 
This avoids evaluating the rational approximation unnecessarily far outside the interval in which its result can be selected, while preserving timing-regular execution.

Accordingly, the implemented \(\tanh\) approximation can be described conceptually as
\begin{equation}
\label{eq:ct_tanh_piecewise}
\tanh(x)
\approx
\begin{cases}
R_{\tanh}(x), & |x| \le \tau_{\tanh},\\
\mathrm{sign}(x), & |x| > \tau_{\tanh},
\end{cases}
\end{equation}
where \(\mathrm{sign}(x)\in\{-1,1\}\) for the saturated cases.

Similarly, the implemented sigmoid approximation is
\begin{equation}
\label{eq:ct_sigmoid_piecewise}
S(x)
\approx
\begin{cases}
\dfrac{1}{2} + \dfrac{1}{2}\,R_{\tanh}\!\left(\dfrac{x}{2}\right), & |x| \le \tau_{S},\\
1, & x > \tau_{S},\\
0, & x < -\tau_{S}.
\end{cases}
\end{equation}

For compactness, define
\[
z_{\mathrm{G}}(x)
=
\sqrt{\frac{2}{\pi}}\left(x+0.044715x^3\right).
\]
The implemented GELU approximation can then be described conceptually as
\begin{equation}
\label{eq:ct_gelu_piecewise}
\mathrm{GELU}(x)
\approx
\begin{cases}
0, & x < -\tau_{\mathrm{GELU}},\\[0.3em]
\dfrac{x}{2}\left(1+R_{\tanh}\!\left(z_{\mathrm{G}}(x)\right)\right),
& |x| \le \tau_{\mathrm{GELU}},\\[0.3em]
x, & x > \tau_{\mathrm{GELU}}.
\end{cases}
\end{equation}

Likewise, the implemented Swish approximation is
\begin{equation}
\label{eq:ct_swish_piecewise}
\mathrm{Swish}(x)
\approx
\begin{cases}
0, & x < -\tau_{\mathrm{Swish}},\\
x\left(
\dfrac{1}{2}
+
\dfrac{1}{2}R_{\tanh}\!\left(\dfrac{x}{2}\right)
\right), & |x| \le \tau_{\mathrm{Swish}},\\
x, & x > \tau_{\mathrm{Swish}}.
\end{cases}
\end{equation}

Equations~(\ref{eq:ct_tanh_piecewise})--(\ref{eq:ct_swish_piecewise}) describe the numerical behavior of the protected implementations. 
In the actual implementation, case selection is not realized through data-dependent branches; instead, the corresponding approximation and saturation candidates are selected using masking-based selection over their bit-level representations. 
Thus, whether the input lies inside or outside the approximation interval does not introduce input-dependent control flow.

\subsection{Selection of Saturation Thresholds}
\label{subsec:threshold}

The saturation thresholds were chosen to balance the approximation error of the \(R_{\tanh}\)-based computation with the error introduced by saturation near the saturation thresholds.
For \(\tanh\), by odd symmetry, it is sufficient to consider the positive saturation threshold. 
According to Equation~(\ref{eq:ct_tanh_piecewise}), two error contributions are relevant near \(x=\tau_{\tanh}\). 
The first is the approximation error of the rational function,
\[
e_{R}(x)=|\tanh(x)-R_{\tanh}(x)|,
\]
while the second is the error introduced by positive saturation,
\[
e_{\mathrm{sat}}(x)=1-\tanh(x).
\]
The threshold \(\tau_{\tanh}\) was selected by balancing these two errors. 
If the threshold is too small, saturation is applied prematurely and the saturation error dominates; 
if it is too large, the rational approximation is used too far into the near-saturated tail and the approximation error dominates.
Accordingly, the threshold was chosen such that
\[
|\tanh(\tau_{\tanh})-R_{\tanh}(\tau_{\tanh})| = 1-\tanh(\tau_{\tanh}).
\]
Solving this equation numerically yields
\[
\tau_{\tanh} \approx 4.97.
\]
The negative saturation threshold follows by symmetry.

The sigmoid threshold is obtained using the same balancing criterion on the positive saturation side. 
We require
\[
\left|
\frac{1}{1+e^{-\tau_S}}
-
\left(
\frac12+\frac12 R_{\tanh}\!\left(\frac{\tau_S}{2}\right)
\right)
\right|
=
1-\frac{1}{1+e^{-\tau_S}}.
\]
Using the identity
\[
\frac{1}{1+e^{-x}}=\frac12+\frac12\tanh\!\left(\frac{x}{2}\right),
\]
this condition simplifies to
\[
\left|
\tanh\!\left(\frac{\tau_S}{2}\right)
-
R_{\tanh}\!\left(\frac{\tau_S}{2}\right)
\right|
=
1-\tanh\!\left(\frac{\tau_S}{2}\right).
\]
This is the same equation as for \(\tanh\), evaluated at \(\tau_S/2\). Therefore,
\[
\tau_S = 2\tau_{\tanh}\approx9.94.
\]
The negative sigmoid threshold follows from the symmetry relation \(S(-x)=1-S(x)\).

For GELU and Swish, the thresholds were selected empirically using the same balancing intuition: the threshold should be large enough to avoid premature saturation, but small enough to prevent the \(R_{\tanh}\)-based computation from being used unnecessarily far into saturated or asymptotic regions.
In the evaluated implementation, we use
\[
\tau_{\mathrm{GELU}} = 3.6
\qquad \text{and} \qquad
\tau_{\mathrm{Swish}} = 8.
\]
For GELU, inputs below \(-\tau_{\mathrm{GELU}}\) are saturated to zero, while inputs above \(\tau_{\mathrm{GELU}}\) are saturated to the linear output \(x\). 
For Swish, inputs below \(-\tau_{\mathrm{Swish}}\) are saturated to zero, while inputs above \(\tau_{\mathrm{Swish}}\) are saturated to \(x\), matching the limiting behavior of \(xS(x)\).

\subsection{Instantiation on ReLU}
\label{subsec:relu_instantiation}

The proposed methodology also applies to simple activation functions such as ReLU, even though they do not require nonlinear approximation.
ReLU is a special case as it does not require dedicated computation as such, its usual implementation,
\[
\mathrm{ReLU}(x)=\max(0,x),
\]
is often compiled into a short and highly efficient instruction sequence that differs substantially from the computation required for the other activation functions. \Cref{code:ReLU_src,code:ReLU_asm} demonstrate this aspect. There is no concrete computation involved, other than a compare (\texttt{vcmpe.f32}). On these few lines of code, the conditional operation (replace value with zero) itself translates into an \texttt{it, vmovlt.f32} pair on any optimization level. Even \texttt{-O0}. While that does not rely on conditional branch instructions, and as such exhibits constant time behavior, its computational profile does differ significantly from those operations actually performing arithmetic instructions.

\begin{lstlisting}[caption={Naive ReLU implementation in C++.}, label=code:ReLU_src]
	float ReLU(float x) {
		return (x > 0.0f) ? x : 0.0f;
	}
\end{lstlisting}
\lstset{language=[x86masm]Assembler}
\begin{lstlisting}[caption={Emitted ARM assembly for \cref{code:ReLU_src} at \texttt{-O3}.}, label=code:ReLU_asm]
ReLU(float):
	vldr.32    s15, .L13 @ tmp117,
	vcmpe.f32  s0, s15   @ x,
	vmrs       APSR_nzcv, FPSCR
	it         lt        @
	vmovlt.f32 s0, s15   @,, tmp117
	bx         lr        @
.L13:
	.word   0
\end{lstlisting}

As a result, directly comparing such an implementation with the approximated nonlinear activations would not provide a meaningful basis for constant-time evaluation across the set of functions considered in this work.

In the present instantiation, this was addressed by augmenting the ReLU computation with a dummy arithmetic component derived from the same \(R_{\tanh}\)-based computation used for the nonlinear activations.
This additional computation does not affect the final ReLU result, but it helps align the computational structure and execution cost with the more complex activation functions. 
Only limited \texttt{NOP} padding was used for fine-grained cycle alignment where necessary.

This design choice avoids relying exclusively on long sequences of \texttt{NOP} instructions, which would provide little resemblance to realistic computation and could produce artificial side-channel characteristics. 
Instead, the implementation preserves a more representative arithmetic profile while still achieving identical timing behavior.

\subsection{Implementation Considerations}
\label{subsec:implementation_considerations}

Several additional measures were taken to preserve the intended constant-time behavior after compilation. 
First, \(R_{\tanh}\) was isolated as a dedicated function in order to control compiler optimization effects: function-scope optimizations are applied or suppressed depending on whether the function is inlined.
Second, synchronization barriers and compiler fences were used around the measurement points to reduce the risk of instruction reordering during benchmarking. 
Additionally, the generated assembly code for these basic building blocks has been inspected to verify that the intended branchless structure was preserved by the compiler and that no unintended input-dependent optimizations were introduced.

Overall, the proposed methodology combines fixed-cost approximation, masking-based saturation logic, masking-based selection, and cycle-alignment techniques to realize activation functions with input-independent execution time on the target embedded platform.

\section{Experimental Setup}
\label{sec:experimental_setup}
This section presents the experimental framework used to evaluate the proposed constant-time implementation methodology as instantiated on the considered activation functions.
Section~\ref{sec:platform} describes the hardware and software platform, Section~\ref{sec:measurement} details the cycle-accurate timing methodology and benchmarking procedure, and Section~\ref{sec:input} defines the evaluated input intervals together with the performance and numerical-accuracy metrics used in the subsequent analysis.

\subsection{Hardware and Software Platform}
\label{sec:platform}
All experiments were performed on an STM32F411E-DISCO development board equipped with an ARM Cortex-M4 microcontroller. 
The processor clock was configured to operate at \(84\)\,MHz using the internal HSI oscillator and PLL-based clock multiplication.

The firmware was implemented in C++ using the GNU++14 language standard (\texttt{-std=gnu++14}) and developed with the STM32 HAL framework. 
The protected activation functions and benchmarking routines were implemented manually, without relying on external machine-learning libraries. 
The unprotected nonlinear baselines used standard single-precision mathematical operations where appropriate. 
The project was compiled in STM32CubeIDE using the GNU Arm Embedded Toolchain (\texttt{arm-none-eabi-g++}) with optimization enabled.

All activation functions were implemented using single-precision floating-point arithmetic (\texttt{float}). 
The Cortex-M4 hardware floating-point unit (FPU) was enabled, and the project was compiled for the FPv4-SP-D16 floating-point architecture using the hardware floating-point ABI (\texttt{-mfloat-abi=hard}).

The program code was executed from flash memory. 
Although the Cortex-M4 does not employ a conventional cache hierarchy, flash prefetch and wait-state effects may still have a minor influence on instruction timing.
The implementation used in this work is publicly available at \url{https://github.com/andrewtyv/activation_timing}.

\subsection{Measurement Methodology}
\label{sec:measurement}
Execution time was measured using the Data Watchpoint and Trace (DWT) cycle counter, which provides cycle-level timing on the target platform. 
For each measurement, a synchronization barrier was executed before sampling the start value of the DWT cycle counter. 
The timed region then included the activation-function call, assignment of the result to a \texttt{volatile} sink variable, and a post-execution synchronization barrier. 
Loop control, input generation, and storage of the recorded cycle count were outside the timed region.

For the grid-based timing benchmarks, each activation function was evaluated independently. 
The three-function setting used the interval \([-8,8]\) with a step size of \(0.01\), whereas the extended five-function setting used the interval \([-500,500]\) with a step size of \(1.0\). 
In both settings, every input value was evaluated five times, and all recorded timing results were retained for analysis without aggregation.

For each benchmark run, one activation function was selected as the measured function, and the same measurement loop was executed for all inputs and repetitions. 
The reported results for different activation functions were obtained from separate runs using the same measurement procedure.

A warm-up execution was performed before the measurement loop in order to reduce the impact of initial flash-access effects. 
To improve the reliability and reproducibility of the measurements, interrupts were disabled during benchmarking. 
In addition, compiler and hardware barriers were placed around the measured region to limit instruction reordering effects. 
In particular, \texttt{atomic\_signal\_fence}, \texttt{\_\_DSB}, and \texttt{\_\_ISB} were used to constrain compiler and processor reordering.

\subsection{Input Range and Evaluation Metrics}
\label{sec:input}
Two input intervals were used in the evaluation. 
The interval \([-8,8]\) was used for the timing and desynchronization experiments in the three-function setting consisting of ReLU, sigmoid, and \(\tanh\), and for the numerical-accuracy evaluation of the approximated sigmoid and \(\tanh\) implementations. 
The extended interval \([-500,500]\) was used to evaluate the constant-time behavior and numerical accuracy of the extended five-function setting consisting of ReLU, sigmoid, \(\tanh\), GELU, and Swish. 
This extended range was chosen to stress the saturated or asymptotic regions of the nonlinear functions and to test whether timing regularity is preserved well beyond the saturation thresholds.

A sampling step of \(0.01\) was used for the interval \([-8,8]\), while a step of \(1.0\) was used for the extended interval \([-500,500]\). 
These sampling steps provide a fine-grained view of timing behavior near the nonlinear and saturation-threshold regions in the three-function setting, and a broad view of timing regularity over the extended range.

For the timing experiments, the primary quantity recorded was the number of CPU cycles per activation-function call. 
For presentation purposes, these measurements were converted to execution time by dividing the measured cycle count by the processor clock frequency of \(84 \times 10^{6}\,\mathrm{Hz}\). 
Presenting the results in time units facilitates comparison of timing characteristics across different activation-function implementations and makes the observed input-independent behavior easier to interpret.

As a secondary criterion, the numerical accuracy of the protected nonlinear activation functions was evaluated against the corresponding reference implementations: the standard sigmoid, \(\tanh\) and GELU functions, as well as Swish with \(\beta=1\).

Approximation quality was quantified using mean squared error (MSE), root mean squared error (RMSE), and maximum absolute error over the sampled input grid used for the corresponding accuracy experiment.
ReLU was excluded from the error tables because it is exact by construction.

Together, these metrics characterize the timing regularity and numerical accuracy of the protected implementations.

\section{Results}
\label{sec:results}

This section evaluates the proposed methodology in four steps. 
Section~\ref{subsec:unprotected_timing} first establishes the timing-leakage baseline by analyzing the unprotected ReLU, sigmoid, and \(\tanh\) implementations. 
Section~\ref{subsec:attack} then examines a desynchronization-based hiding countermeasure and shows that, despite visually obscuring the timing traces, it remains vulnerable to a profiled template-based timing attack. 
Section~\ref{subsec:ct_three_functions} evaluates the proposed constant-time implementations in the three-function setting, including both timing regularity and numerical accuracy over the interval \([-8,8]\). 
Finally, Section~\ref{subsec:extended_activation_set} extends the evaluation to GELU and Swish over the wider interval \([-500,500]\), demonstrating that the proposed methodology generalizes to an extended five-function setting.

\begin{figure}[htb]
    \centering
        \includegraphics[width=0.8\linewidth]{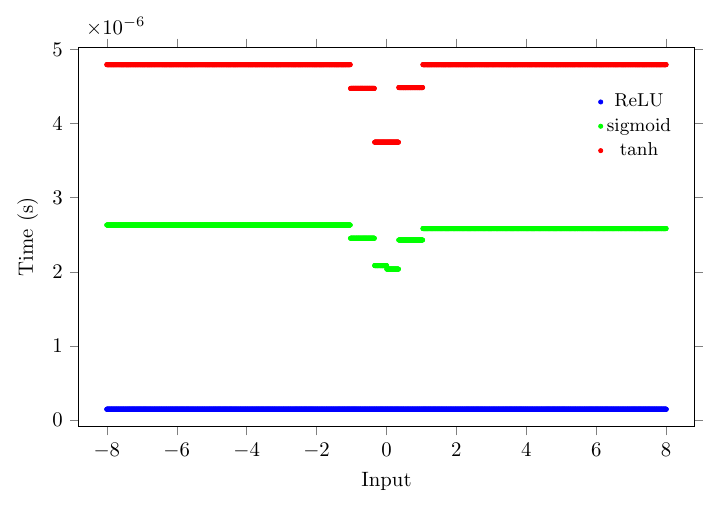}
    \caption{Execution time as a function of input for the unprotected activation-function implementations. The timing profiles are clearly distinguishable across functions, and sigmoid and \(\tanh\) exhibit visible input-dependent timing variation.}
    \label{fig:activation_timing}
\end{figure}

\subsection{Timing Leakage in Unprotected Activation Functions}
\label{subsec:unprotected_timing}

Figure~\ref{fig:activation_timing} shows the timing behavior of the unprotected implementations. 
Clear differences can be observed both across activation functions and, for the nonlinear activations, across input values. ReLU exhibits the lowest execution time and remains close to \(12\) cycles (\(0.143\,\mu\text{s}\)) over the evaluated domain, whereas the unprotected sigmoid and \(\tanh\) implementations require substantially longer execution times, approximately \(221\) cycles (\(2.63\,\mu\text{s}\)) and \(403\) cycles (\(4.80\,\mu\text{s}\)), respectively.
In addition, sigmoid and \(\tanh\) show visible timing variations in the central input region, producing activation-specific timing signatures that can be distinguished from one another. 
These observations establish a clear timing-leakage baseline for the unprotected implementations and are consistent with previously reported timing side-channel leakage in neural-network computations~\cite{batina2019csi,breier2023desynchronization}.

\subsection{Template-Based Attack on Desynchronized Implementations}
\label{subsec:attack}

\begin{figure}[htb]
    \centering
    \includegraphics[width=0.9\linewidth]{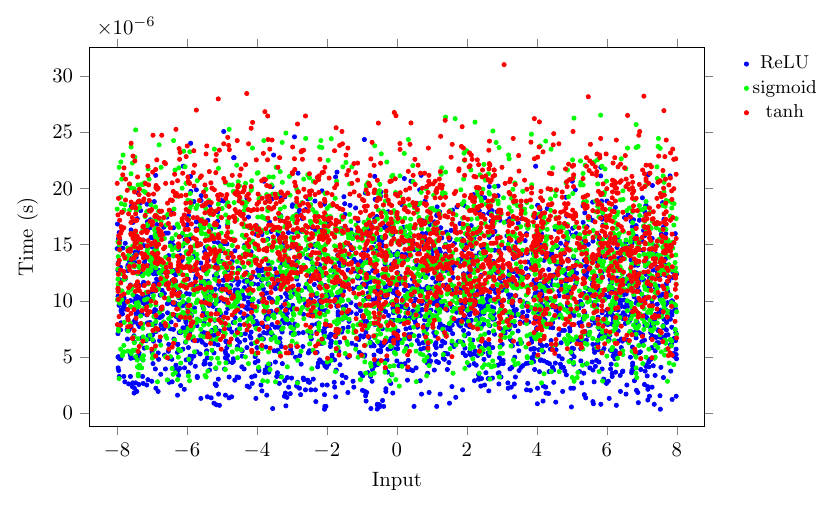}
    \caption{Execution time after applying the desynchronization-based countermeasure~\cite{breier2023desynchronization}. The original timing patterns are no longer distinguishable by visual inspection.}
    \label{fig:jitter}
\end{figure}

Having established that the unprotected implementations exhibit distinguishable timing behavior, we next examine whether hiding this behavior through desynchronization is sufficient to prevent activation-function identification from timing measurements.
For this purpose, we implemented a desynchronization-style countermeasure inspired by~\cite{breier2023desynchronization}, in which a nonnegative random delay, sampled from the distribution used in the experiment, is added to each activation-function evaluation to obscure the original timing pattern.

The execution time of the unprotected ReLU, sigmoid, and \(\tanh\) implementations was first recorded for \(8{,}000\) random inputs in the interval \([-8,8]\) for each activation function. Based on these measurements, the parameters of the added delay were selected so that the resulting timing behavior becomes visibly desynchronized across the three activation functions. The resulting execution times are shown in Figure~\ref{fig:jitter}. 
After randomization, the original timing structure is visually obscured, and the three activation functions can no longer be reliably distinguished by inspection alone.

\begin{figure*}[htb]
    \centering

    \begin{subfigure}{0.48\linewidth}
        \centering
        \includegraphics[width=\linewidth]{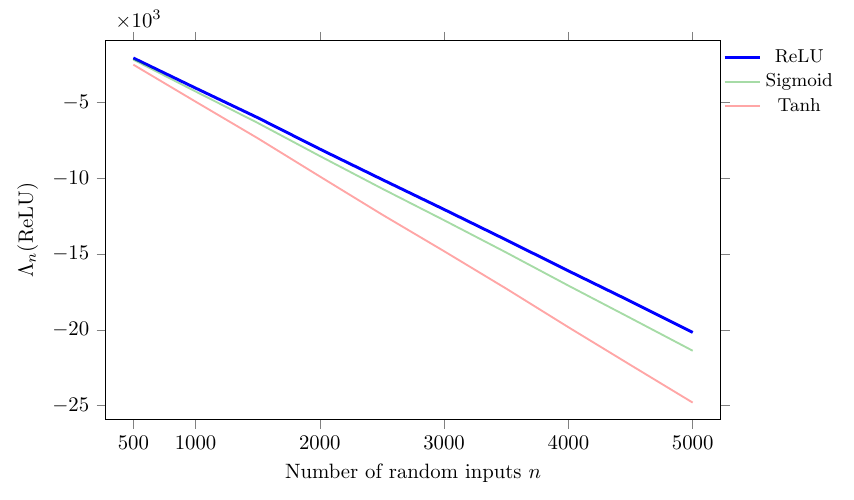}
        \caption{True activation: ReLU}
        \label{fig:relu_attack}
    \end{subfigure}
    \hfill
    \begin{subfigure}{0.48\linewidth}
        \centering
        \includegraphics[width=\linewidth]{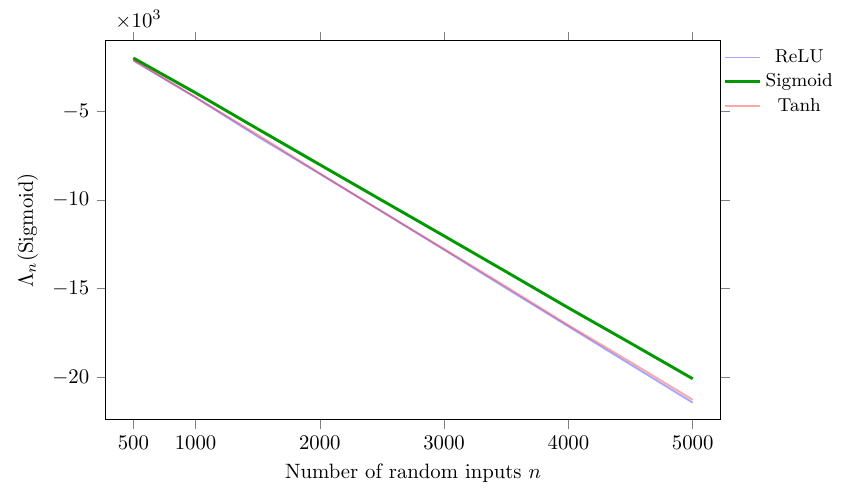}
        \caption{True activation: sigmoid}
        \label{fig:sigmoid_attack}
    \end{subfigure}

    \vspace{0.8em}

    \begin{subfigure}{0.48\linewidth}
        \centering
        \includegraphics[width=\linewidth]{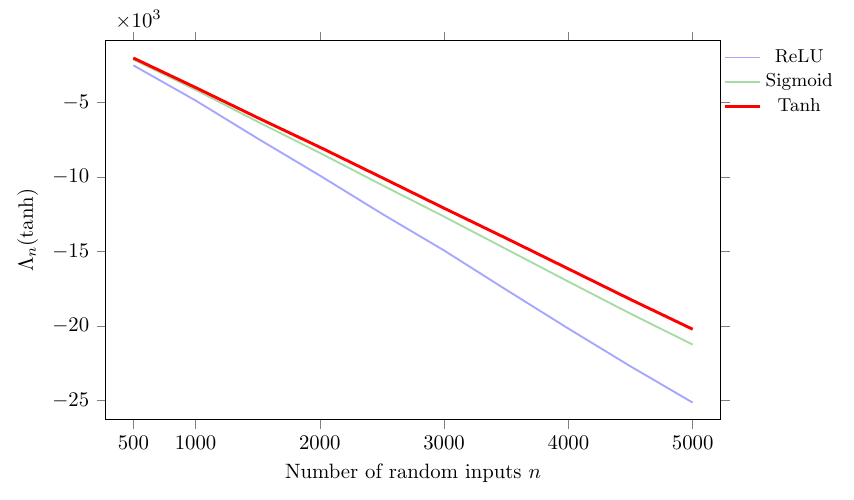}
        \caption{True activation: \(\tanh\)}
        \label{fig:tanh_attack}
    \end{subfigure}

    \caption{
    Evolution of the accumulated class scores under the template-based attack for the three possible true activation functions. In each panel, the true activation function is fixed, while the three curves correspond to the competing activation-function hypotheses.}
    \label{fig:desync_attack}
\end{figure*}

To evaluate whether this protection can be bypassed, we adopt a profiled template-based attack, following the standard methodology commonly used in power and electromagnetic side-channel analysis of cryptographic implementations~\cite[Section~4.3]{hou2024cryptography}. For each class \(c \in \{\mathrm{ReLU}, \mathrm{sigmoid}, \tanh\}\), we collected \(N_{\mathrm{prof}} = 10{,}000\) profiling measurements. 
Each measurement consisted of the execution time of the desynchronized activation-function implementation for a random input from the evaluated domain.
From these measurements, a univariate Gaussian template was estimated for each class:
\[
\mu_c = \frac{1}{N_{\mathrm{prof}}} \sum_{i=1}^{N_{\mathrm{prof}}} t_i^{(c)},
\qquad
\sigma_c^2 = \frac{1}{N_{\mathrm{prof}}-1} \sum_{i=1}^{N_{\mathrm{prof}}} \left(t_i^{(c)} - \mu_c\right)^2,
\]
where \(t_i^{(c)}\) denotes the \(i\)-th measured execution time for class \(c\).

The resulting template parameters were
\[
(\mu_{\mathrm{ReLU}}, \sigma_{\mathrm{ReLU}}^2) = (9.964\,\mu\mathrm{s},\,20.346\,\mu\mathrm{s}^2),
\]
\[
(\mu_{\mathrm{sigmoid}}, \sigma_{\mathrm{sigmoid}}^2) = (12.380\,\mu\mathrm{s},\,20.575\,\mu\mathrm{s}^2),
\]
\[
(\mu_{\tanh}, \sigma_{\tanh}^2) = (14.520\,\mu\mathrm{s},\,20.645\,\mu\mathrm{s}^2).
\]

During the attack phase, the true activation function was fixed but unknown to the attacker. 
Let \(t_1, t_2, \dots, t_n\) denote the execution times obtained for \(n\) random inputs in the interval \([-8,8]\). 
Following standard template-based attack methodology~\cite[Section~4.3]{hou2024cryptography}, and assuming independence between measurements, the likelihood of class \(c\) after \(n\) observations is given by
\[
\mathcal{L}_n(c)
=
\prod_{i=1}^{n}
\frac{1}{\sqrt{2\pi\sigma_c^2}}
\exp\left(
-\frac{(t_i-\mu_c)^2}{2\sigma_c^2}
\right).
\]
Equivalently, and more conveniently for numerical evaluation, the accumulated class score can be written as
\[
\Lambda_n(c)
=
-\sum_{i=1}^{n}
\left(
\ln \sigma_c^2 + \frac{(t_i-\mu_c)^2}{\sigma_c^2}
\right),
\]
where the common additive term involving \(\ln(2\pi)\) is omitted and the common positive factor \(1/2\) is absorbed into the score scaling, since neither affects the comparison between classes.
As the number of observations increases, the score associated with the correct class is expected to separate progressively from the competing hypotheses.

The attack results are shown in Figure~\ref{fig:desync_attack}, which plots the evolution of the accumulated class scores as the number of measurements increases, with each of the three activation functions considered in turn as the true class. 
In all three cases, the score of the correct class separates from the competing hypotheses after approximately \(2{,}000\) measurements.

These results show that desynchronization can obscure the timing pattern of individual measurements, but it does not eliminate the class-dependent statistical structure of the execution-time distributions. As additional measurements are collected, the correct activation function becomes increasingly favored under the corresponding template, enabling a profiled adversary to identify the activation class with growing confidence. This directly motivates the constant-time design proposed in this work, whose objective is to remove the input-dependent timing behavior itself rather than merely randomize its temporal position.

\subsection{Constant-Time Evaluation of ReLU, Sigmoid, and \(\tanh\)}
\label{subsec:ct_three_functions}
Next, we evaluate the proposed constant-time implementations in the three-function setting consisting of ReLU, sigmoid, and \(\tanh\). 
The evaluation considers timing regularity for all three functions and numerical accuracy for the two approximated nonlinear functions over the interval \([-8,8]\).

\begin{figure*}[tb]
    \centering

    \begin{subfigure}{0.4\linewidth}
        \centering
        \includegraphics[width=\linewidth]{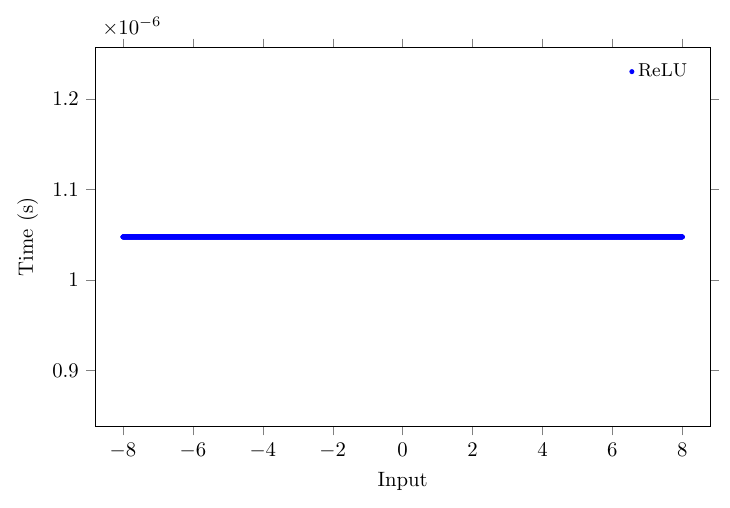}
        \caption{ReLU}
    \end{subfigure}
    \hfill
    \begin{subfigure}{0.4\linewidth}
        \centering
        \includegraphics[width=\linewidth]{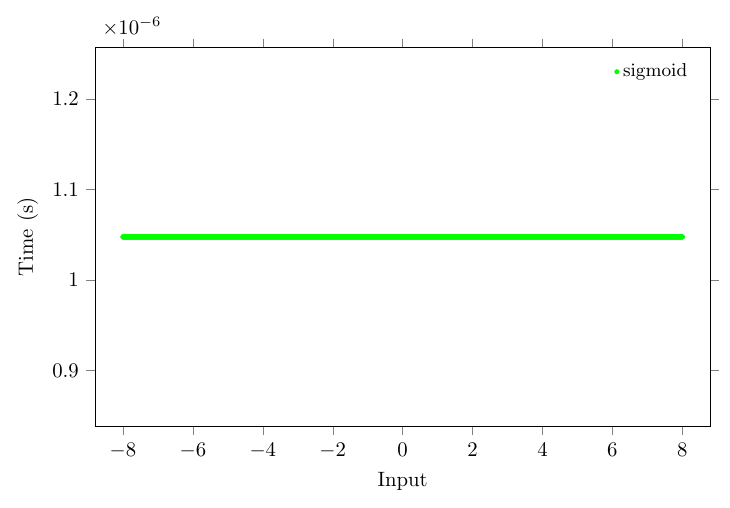}
        \caption{sigmoid}
    \end{subfigure}

    \begin{subfigure}{0.4\linewidth}
        \centering
        \includegraphics[width=\linewidth]{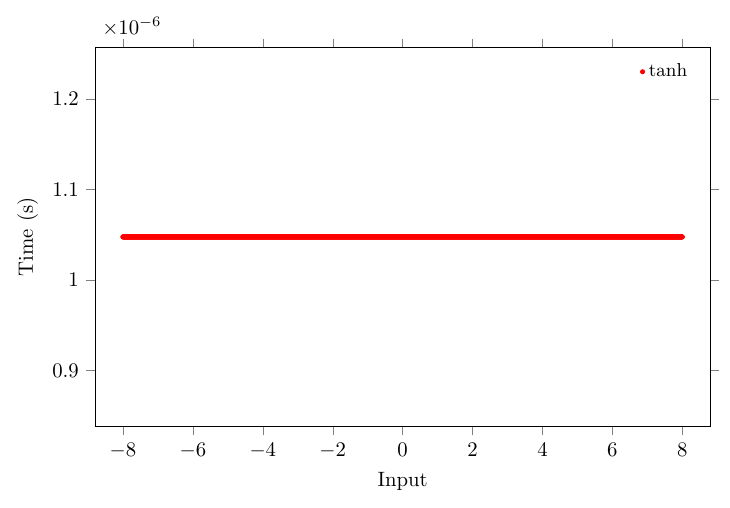}
        \caption{\(\tanh\)}
    \end{subfigure}
    \hfill
    \begin{subfigure}{0.4\linewidth}
        \centering
        \includegraphics[width=\linewidth]{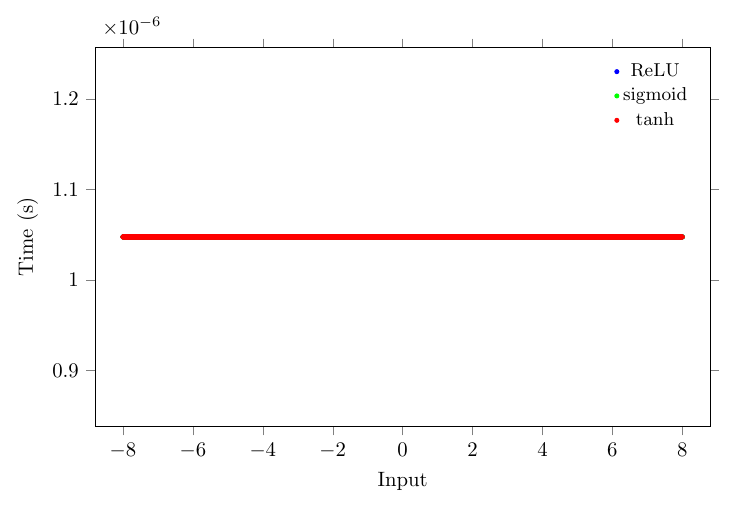}
        \caption{relative comparison}
    \end{subfigure}
    \caption{Execution time as a function of input for the protected activation-function implementations: (a) ReLU, (b) sigmoid, (c) \(\tanh\), and (d) relative comparison. 
    All three implementations exhibit constant execution time over the evaluated input range and are fully aligned at \(88\) cycles (\(1.048\,\mu\text{s}\)).}
    \label{fig:activation_timing_protected}
\end{figure*}

The protected implementations shown in Figure~\ref{fig:activation_timing_protected} exhibit flat timing profiles throughout the evaluated input range. 
For ReLU, sigmoid, and \(\tanh\), the measured execution time remains constant at \(1.048\,\mu\text{s}\), corresponding to \(88\) clock cycles.
No input-dependent timing variation is observable in the individual plots, and the relative comparison in Figure~\ref{fig:activation_timing_protected}(d) shows that the three protected implementations are fully aligned in execution time. 
These results confirm that the proposed methodology suppresses the timing leakage present in the unprotected baseline for the considered activation functions.

\paragraph{Numerical accuracy.}

In addition to timing behavior, we evaluated the numerical accuracy of the approximated sigmoid and \(\tanh\) implementations over the interval \([-8,8]\) by comparison with the corresponding reference implementations. 
Since the ReLU implementation is exact by construction, approximation error was assessed only for sigmoid and \(\tanh\). 

\begin{table}[htb]
\centering
\caption{Numerical accuracy of the approximated activation functions over the input interval $[-8,8]$.}
\label{tab:activation_error}
\begin{tabular}{lccc}
\hline
Activation & MSE & RMSE & Max.\ abs. error \\
\hline
Sigmoid & $2.93 \times 10^{-12}$ & $1.71 \times 10^{-6}$ & $7.51 \times 10^{-6}$ \\
$\tanh$ & $6.14 \times 10^{-10}$ & $2.48 \times 10^{-5}$ & $9.59 \times 10^{-5}$ \\
\hline
\end{tabular}
\end{table}

\begin{figure}[htb]
\centering
\includegraphics[width=0.9\columnwidth]{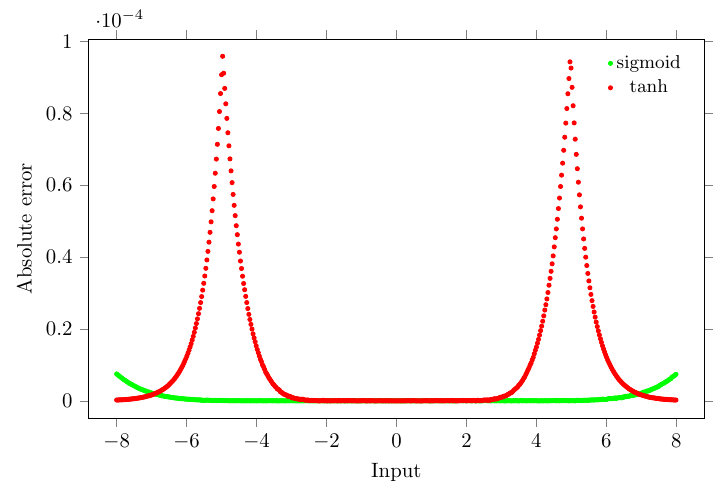}
\caption{Absolute error of the approximated sigmoid and \(\tanh\) functions over the evaluated input interval.}
\label{fig:absolute_error}
\end{figure}

Table~\ref{tab:activation_error} summarizes the obtained error metrics. 
The sigmoid approximation achieved very high accuracy, with an MSE of $2.93 \times 10^{-12}$, an RMSE of $1.71 \times 10^{-6}$, and a maximum absolute error of $7.51 \times 10^{-6}$. 
The \(\tanh\) approximation exhibited slightly larger deviation, with an MSE of $6.14 \times 10^{-10}$, an RMSE of $2.48 \times 10^{-5}$, and a maximum absolute error of $9.59 \times 10^{-5}$.

Figure~\ref{fig:absolute_error} shows the absolute error as a function of the input value. 
For sigmoid, the error remains very small over the full evaluated domain and increases only slightly toward the interval boundaries. 
For \(\tanh\), the largest error occurs around the saturation thresholds \(\pm\tau_{\tanh}\), where the implementation switches from a rational approximation to saturated output values. 
Away from the saturation thresholds, the error rapidly decreases and remains negligible. 

Overall, both approximations remain close to the corresponding reference functions in the evaluated domain.

\paragraph{Execution-Time Cost}
Applying the proposed methodology introduces different execution-time effects depending on the activation function.
For ReLU, the protected implementation is slower than the unprotected baseline, increasing from approximately \(12\) cycles (\(0.143\,\mu\text{s}\)) to \(88\) cycles (\(1.048\,\mu\text{s}\)). 
This increase is expected, since the original ReLU computation is extremely simple, whereas the protected implementation was intentionally augmented to match the fixed computational structure required for timing alignment. 

By contrast, the protected sigmoid and \(\tanh\) implementations are faster than their unprotected counterparts. 
In the evaluated interval \([-8,8]\), the unprotected sigmoid is executed in approximately \(221\) cycles (\(2.63\,\mu\text{s}\)), whereas the protected version requires only \(88\) cycles (\(1.048\,\mu\text{s}\)). Similarly, the unprotected \(\tanh\) runs in approximately \(403\) cycles (\(4.80\,\mu\text{s}\)), while the protected implementation also requires \(88\) cycles (\(1.048\,\mu\text{s}\)).
This reduction is explained by the use of fixed-cost rational approximations (Equations~\ref{eq:core} and~\ref{eq:sigmoid_core}) in place of standard-library exponential computations, which are more expensive on the target platform. 

\subsection{Extended Constant-Time Evaluation Across Five Activation Functions}
\label{subsec:extended_activation_set}

To further evaluate the generality of the proposed methodology, we extended the set of protected implementations to include GELU and Swish, in addition to ReLU, sigmoid, and \(\tanh\). 

\begin{figure}[htb]
    \centering
    \includegraphics[width=0.85\linewidth]{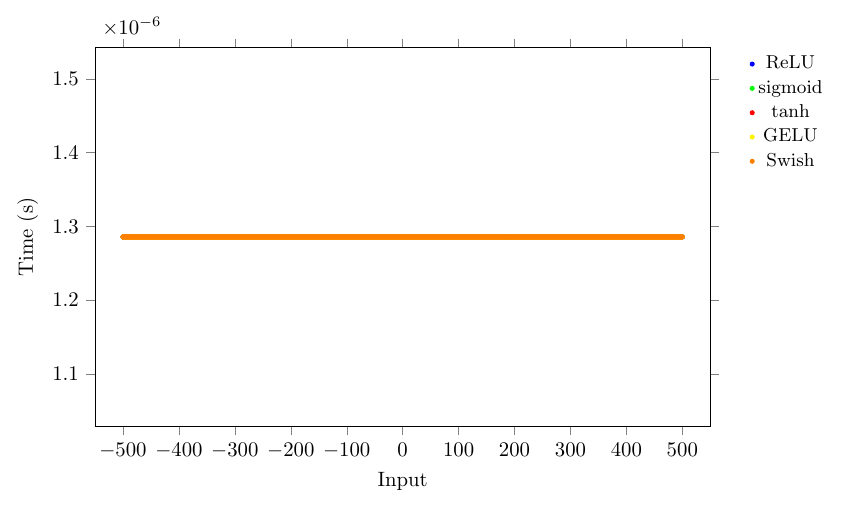}
    \caption{Execution time as a function of input for the extended set of protected activation-function implementations over the interval \([-500,500]\). 
    ReLU, sigmoid, \(\tanh\), GELU, and Swish all exhibit identical execution latency of \(108\) cycles, corresponding to approximately \(1.286\,\mu\mathrm{s}\) at \(84\)\,MHz.}
    \label{fig:five_functions_time}
\end{figure}

The extended evaluation was performed over the input interval \([-500,500]\). 
Across this range, all five protected activation functions exhibited identical execution latency of \(108\) clock cycles, corresponding to approximately \(1.286\,\mu\mathrm{s}\) at the \(84\)\,MHz clock frequency of the target platform. 
Figure~\ref{fig:five_functions_time} shows that the timing profiles of ReLU, sigmoid, \(\tanh\), GELU, and Swish are fully aligned throughout the extended input range.

This result indicates that the proposed masking-based selection, fixed-cost Pad\'e approximation, and cycle-alignment strategy can be applied consistently beyond the original set of activation functions.

\begin{figure}[htb]
    \centering
    \includegraphics[width=0.9\linewidth]{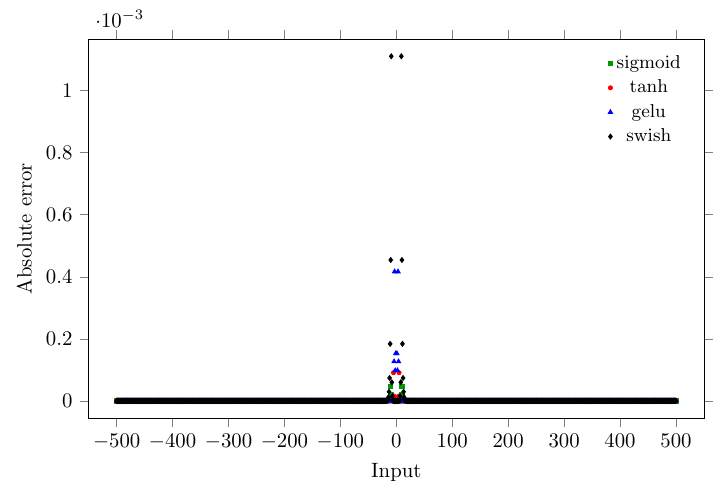}
    \caption{Absolute error of the protected sigmoid, \(\tanh\), GELU, and Swish implementations over the extended input interval \([-500,500]\). 
    The errors are concentrated around the saturation thresholds, while they remain negligible over almost the entire evaluated range.}
    \label{fig:extended_absolute_error}
\end{figure}

\paragraph{Numerical accuracy.}
The numerical accuracy of the extended protected implementations was also evaluated over the interval \([-500,500]\). 
Since ReLU remains exact by construction, Table~\ref{tab:extended_activation_error} reports the error metrics only for the four nonlinear activation functions. 
Figure~\ref{fig:extended_absolute_error} shows the absolute error as a function of the input value for the four nonlinear protected activation functions. 

\begin{table}[htb]
\centering
\caption{Numerical accuracy of the protected nonlinear activation functions over the input interval \([-500,500]\).}
\label{tab:extended_activation_error}
\begin{tabular}{lccc}
\hline
Activation & MSE & RMSE & Max.\ abs. error \\
\hline
Sigmoid & $5.78 \times 10^{-12}$ & $2.40 \times 10^{-6}$ & $4.54 \times 10^{-5}$ \\
$\tanh$ & $1.72 \times 10^{-11}$ & $4.15 \times 10^{-6}$ & $9.08 \times 10^{-5}$ \\
GELU & $4.45 \times 10^{-10}$ & $2.11 \times 10^{-5}$ & $4.17 \times 10^{-4}$ \\
Swish & $2.96 \times 10^{-9}$ & $5.44 \times 10^{-5}$ & $1.11 \times 10^{-3}$ \\
\hline
\end{tabular}
\end{table}

The error is highly localized: away from the saturation thresholds, the curves remain essentially flat and close to zero in the extended interval \([-500,500]\).
This behavior is especially visible for sigmoid and \(\tanh\), whose errors become negligible over the extended range because both functions reach their saturated regions over most of the evaluated domain.
Consequently, although the evaluation interval is substantially larger than the original interval \([-8,8]\), the aggregate error metrics for sigmoid and \(\tanh\) remain very small.

GELU exhibits low error, with a maximum absolute error of $4.17 \times 10^{-4}$. Although roughly an order of magnitude larger than the sigmoid and \(\tanh\) maxima, this error remains localized near the saturation threshold and is negligible in most of the evaluated domain.
Swish has the largest deviation among the evaluated nonlinear functions, reaching a maximum absolute error of $1.11 \times 10^{-3}$. 
This is expected because Swish multiplies the sigmoid component by the input value \(x\), so small approximation or saturation errors in the sigmoid component can be amplified around the saturation thresholds.
Nevertheless, the Swish error is still confined to a narrow input region, while the error remains negligible over the vast majority of the extended domain.

Overall, the timing and accuracy results show that the contribution of the present work is not limited to ReLU, sigmoid, and \(\tanh\), but extends to a broader methodology for constructing constant-time activation functions on embedded platforms.

\paragraph{Execution-Time Cost}

\begin{table}[htb]
\centering
\caption{Execution-time statistics of the original unprotected activation-function implementations in the five-function setting, reported in \(\mu\mathrm{s}\).}
\label{tab:unprotected_five_function_timing}
\begin{tabular}{lccccc}
\hline
Activation & Min. & Mean & Median & Std. & Max. \\
\hline
ReLU       & \(0.107\) & \(0.107\) & \(0.107\) & \(0.000\) & \(0.107\) \\
Sigmoid    & \(1.440\) & \(1.921\) & \(1.893\) & \(0.273\) & \(2.524\) \\
\(\tanh\)  & \(1.452\) & \(1.871\) & \(1.905\) & \(0.214\) & \(2.679\) \\
GELU       & \(1.143\) & \(1.232\) & \(1.167\) & \(0.540\) & \(5.988\) \\
Swish      & \(1.595\) & \(2.076\) & \(2.048\) & \(0.273\) & \(2.798\) \\
\hline
\end{tabular}
\end{table}

The extended set of implementations increases the common protected latency from \(88\) cycles in the three-function setting to \(108\) cycles in the extended five-function setting, corresponding to approximately \(1.286\,\mu\mathrm{s}\) at \(84\)\,MHz. 
This increase reflects the additional fixed-cost arithmetic and cycle alignment required to support GELU and Swish within a common execution budget.
Table~\ref{tab:unprotected_five_function_timing} summarizes the execution-time statistics of the original unprotected implementations used for this comparison.

To put this cost into context, the original unprotected implementations in the five-function setting exhibited substantially different execution-time characteristics. 
Over \(5005\) measurements per function, the unprotected ReLU implementation executed in \(9\) cycles (\(0.107\,\mu\mathrm{s}\)), while the unprotected sigmoid, \(\tanh\), GELU, and Swish implementations had mean execution times of \(161\) cycles (\(1.921\,\mu\mathrm{s}\)), \(157\) cycles (\(1.871\,\mu\mathrm{s}\)), \(103\) cycles (\(1.232\,\mu\mathrm{s}\)), and \(174\) cycles (\(2.076\,\mu\mathrm{s}\)), respectively.

Thus, the protected implementation introduces a substantial overhead for ReLU, increasing its latency from \(9\) cycles (\(0.107\,\mu\mathrm{s}\)) to \(108\) cycles (\(1.286\,\mu\mathrm{s}\)).
This is expected, because ReLU is inherently much simpler than the nonlinear activations and must be augmented with dummy arithmetic to align its timing with the common protected execution budget. 
In contrast, the protected implementations are faster on average than the original sigmoid, \(\tanh\), and Swish implementations, reducing their mean execution times by approximately \(33.1\%\), \(31.3\%\), and \(38.1\%\), respectively. 
For GELU, the protected latency is only slightly higher than the original mean latency, increasing from \(103\) cycles (\(1.232\,\mu\mathrm{s}\)) to \(108\) cycles (\(1.286\,\mu\mathrm{s}\)), while also removing the large timing spread observed in the unprotected implementation, whose maximum measured latency reached \(503\) cycles (\(5.988\,\mu\mathrm{s}\)).

Overall, the extended protected implementation trades the extremely low cost of ReLU for a fixed, input-independent execution budget shared by all five activation functions. 
From the perspective of embedded inference, the resulting \(108\)-cycle (\(1.286\,\mu\mathrm{s}\)) latency remains modest, while simultaneously eliminating the activation-dependent and input-dependent timing variation observed in the original implementations.

\section{Discussion}
\label{sec:discussion}
\subsection{Security Implications}

Activation-function leakage has already been shown to support reverse engineering and model analysis~\cite{batina2019csi,horvath2024sok}. 
In this light, protecting activation functions against timing side-channel attacks is particularly relevant for privacy-sensitive and security-critical edge deployments, where inference is executed locally on microcontrollers and timing information may be observable to an attacker. 
Representative examples include wearable and medical sensing devices, as well as other edge-AI systems deployed in adversarial or untrusted environments, where model behavior, structure, or processed inputs may carry security or privacy value~\cite{rocha2024edge,xi2025integrating}.

At the same time, constant-time activations represent only one element of a side-channel-resistant inference pipeline. 
Other parts of the implementation may still introduce exploitable leakage if they are not designed with similar care. 
The proposed approach should therefore be viewed as a building block toward more comprehensive secure embedded inference rather than as a complete countermeasure on its own.

\subsection{Limitations and Future Work}

In the considered setting, the selected rational approximation and saturation thresholds provide a practical compromise between timing regularity, computational cost, and numerical precision. 
However, portability should be interpreted with care, since exact cycle counts and timing-regular code generation may still depend on the target architecture, compiler, and optimization settings.

In addition, the main focus of this work is timing leakage. 
Other side channels, such as power consumption or electromagnetic emanations, were not studied in the same depth. 
Finally, the analysis was performed at the level of individual activation functions rather than complete neural-network inference. 
Future work should therefore extend the study to additional hardware platforms, additional activation families, and end-to-end network implementations.

More broadly, the results indicate that the contribution of this work lies not only in the specific protected implementations of ReLU, sigmoid, \(\tanh\), GELU, and Swish, but in the underlying implementation methodology that combines fixed-cost Pad\'e-based approximation, masking-based selection, dummy arithmetic where needed, and cycle alignment for activation-function evaluation.

\section{Conclusion}
\label{sec:conclusion}

This work proposed a constant-time implementation methodology for activation functions on embedded microcontrollers and validated it on ReLU, sigmoid, \(\tanh\), GELU, and Swish for embedded neural-network inference on a Cortex-M4 platform.
The results showed that unprotected activation functions exhibit distinguishable timing behavior, and that desynchronization-based hiding alone does not fully prevent statistical discrimination under a template-based attack. 
To address this problem, we proposed a methodology based on masking-based selection, cycle alignment, and, for nonlinear activations, a shared rational approximation.

Experimental evaluation demonstrated that the resulting protected implementations achieve identical cycle counts for all tested inputs, including a common latency of \(108\) cycles for all five activation functions.
At the same time, the numerical-accuracy results showed that the approximations remain accurate over the considered input ranges, with sigmoid and \(\tanh\) errors becoming negligible over most of the considered interval and GELU and Swish retaining low error localized around the saturation thresholds.

Overall, the proposed methodology shows that side-channel-aware activation-function design can be achieved in practice with modest implementation cost and low numerical error, making it a promising building block for more secure embedded neural-network inference.

\section*{Acknowledgment}
This work was supported by the Open Research Fund of The State Key Laboratory of Blockchain and Data Security (Grant number: A2566), Zhejiang University.

The authors used ChatGPT (OpenAI) and Claude (Anthropic) for language polishing and editorial assistance.
All generated suggestions were reviewed, revised where necessary, and verified by the authors, who take full responsibility for the final content.

\balance
\bibliographystyle{IEEEtran}
\bibliography{bib}

\end{document}